\documentclass[aps,pre,amsmath,amsfonts,floatfix,superscriptaddress,twocolumn,nofootinbib]{revtex4}
\usepackage{mathrsfs} \usepackage{graphicx} \usepackage{bbm}

\def\AA{\mathring{\text{A}}}

\let\a=\alpha \let\b=\beta \let\g=\gamma \let\d=\delta
\let\e=\varepsilon   
    
\let\s=\sigma \let\t=\tau \let\f=\varphi 
   
\let\D=\Delta   
   
 \let\r=\rho  \let\io=\infty

\def\ie{{i.e. }}\def\eg{{e.g. }}

\def\NN{{\cal N}} 
\def\RR{{\cal R}}

\def\to{\rightarrow}

\newcommand{\beq}{\begin{equation}} \newcommand{\eeq}{\end{equation}}
 \newcommand{\wt}{\widetilde}

\begin{document}

\title{
Leggett's bound for amorphous solids
} 

\author{Giulio Biroli}
\affiliation{Institut de Physique Th{\'e}orique (IPhT),
CEA, and CNRS URA 2306, F-91191 Gif-sur-Yvette, France}

\author{Bryan Clark}
\affiliation{Princeton Center for
  Theoretical Science, Princeton University, Princeton, NJ 08544, USA}

\author{Laura Foini}
\affiliation{LPTENS, CNRS UMR 8549,
  associ\'ee \`a l'UPMC Paris 06, 24 Rue Lhomond, 75005 Paris,
  France.}
\affiliation{
SISSA and INFN, Sezione di Trieste, via Bonomea 265, I-34136 Trieste, Italy
}

\author{Francesco Zamponi}
\affiliation{LPTENS, CNRS UMR 8549, associ\'ee \`a l'UPMC Paris 06, 24
  Rue Lhomond, 75005 Paris, France.}

\begin{abstract}
We investigate the constraints on the superfluid fraction of an amorphous solid following from an upper bound derived by Leggett.  In order to accomplish this, we use as input density profiles generated for amorphous solids in a variety of different manners including by investigating Gaussian fluctuations around classical results.  These rough estimates suggest that, at least at the level of the upper bound, there is not much difference in terms of superfluidity between a glass and a crystal characterized by the same Lindemann ratio. Moreover, we perform Path Integral Monte Carlo simulations of distinguishable Helium 4 rapidly quenched from the liquid phase to very lower temperature, at the density of the freezing transition.  We find that the system crystallizes very quickly, without any sign of intermediate glassiness. Overall our results suggest that the experimental observations of large superfluid fractions in Helium 4 particles after a rapid quench correspond to samples evolving far from equilibrium, instead of being in a stable glass phase. 
Other scenarios and comparisons to other results on the super-glass phase are also discussed.    
\end{abstract}

\maketitle

\section{Introduction}

Recent experiments on solid He$^4$ by Kim and Chan~\cite{Chan1,Chan2,balibarreview}
raised, among many others, the important question of whether disorder can foster
the formation of superfluidity in solid samples.
Following earlier theoretical analyses~\cite{dis1,dis2,BPS1},
Ritner and Reppy~\cite{reppy1,reppy2}
showed that fast quenches produce disordered samples with
a change in the moment of inertia that corresponds to an
extremely high fraction of superfluid density, on the order of $20\%$.
In addition, the role of He$^3$ impurities~\cite{balibarreview}
suggest that disorder must play an important role in the experiments.
Other studies~\cite{norho_s} suggest that the role of disorder is not
to enhance the superfluid fraction but instead to induce non-equilibrium states
in the sample that modify the moment of inertia as a function of temperature and
frequency.  Consequently, in spite of a long series of theoretical and experimental
studies, the relationship between disorder and superfluidity in quantum solids is still
not clear.

Here we want to focus on one particular proposal that was put forward by
Boninsegni et al.~\cite{BPS1}: the possibility of a {\it bulk long-lived metastable
glass phase} of He$^4$.
These authors
performed a Path Integral Monte Carlo (PIMC) numerical simulation of Helium 4
at relatively high density ($\r \sim 0.03$ $\AA$$^{-3}$), where the system was
very quickly quenched from the equilibrium liquid phase at high $T$ to a low temperature
$T=0.2$ K, at which the HCP solid phase is stable. They reported the observation of a phase 
which is structurally similar to the liquid, and with a fraction of superfluid density as high as $60\%$;
this phase was observed to last for a large number of Monte Carlo sweeps before the system
eventually freezes into the equilibrium ordered solid. Boninsegni et al. labeled this the 
``superglass'' phase. Actually, the experimental protocols used to solidify Helium likely produce very disordered solids, 
possibly glasses. In fact the experiments in~\cite{Hunt09} showed evidences of very slow dynamics, the hallmark 
of glassy behavior. The natural and still open question is why freezing in an amorphous density profile should 
enhance superfluidity compared to the crystalline case, which instead is thought to show 
zero or very small condensate fractions \cite{crystal1,crystal2}. Superfluidity is related to exchange, 
which is a local process and
depends mostly on the local neighborhood of a particle. 
Thus, one might expect, contrary to the findings discussed above, 
that dense glasses should have a fraction of superfluid density comparable to the one of 
crystals at the same particle density.
Indeed, a theoretical investigation of the superglass
phase in a simplified (and yet realistic) model of interacting bosons found an extremely small
condensate fraction in the superglass phase~\cite{BCZ08}.
Clearly, the relation between disorder and superfluidity deserves further investigation,
in order to reach a better microscopic understanding of superfluidity in amorphous solids and to
explain the numerical and experimental results.

The main difficulty in the numerical investigation of this problem comes from the fact that
the glass phase (if any) is always expected to be metastable with respect to the crystal phase, which
is the true equilibrium phase of solid Helium.
In a classical system, it is reasonably straightforward to get properties of a metastable phase 
or a glass, because one can easily simulate the physical dynamics of the system by solving Newton's equations
of motion~\cite{KA95}.
In contrast, the real-time dynamics of quantum systems is not accessible numerically because
of the sign problem, and calculating properties involving glassy quantum system is problematic.  
Previous numerical work of Boninsegni et al.~\cite{BPS1} has looked at the fraction of superfluid density 
of a quenched Helium 4 via directly calculating it for a system whose 
PIMC dynamics slowly equilibrates. More recently, a quantum version of the Mode-Coupling Theory
of dynamics in glasses has been developed and compared with Path Integral Molecular Dynamics (PIMD)
simulations~\cite{QMCT}, obtaining accurate informations on the glass transition in quantum hard spheres. 
However, in this study exchange effects were neglected and therefore superfluidity could not be investigated.
Therefore, for the moment path integral simulations are not conclusive.

Here we approach the problem in a different way.
In one of the first works on supersolidity, Leggett showed how one can derive an upper bound for the 
fraction of superfluid density 
of a generic many-body system in which translational invariance is broken,
by means of a variational computation~\cite{Le70}. The output of Leggett's computation is a formula that needs
as only input the average density profile of the solid. This formula has been applied to Helium
crystals, and the aim of this work is to use it to study the amorphous solid.
At present, there is not yet any reliable first principle computation or experimental measurements
of the density profile of amorphous Helium 4.
We endeavor to generate robust estimates of it using a number of different techniques,
in particular by investigating a model of zero-point Gaussian fluctuations around classical configurations,
and PIMC simulations without exchange (which should be closer to the classical dynamics).
Checking whether these techniques all give roughly similar orders for the bound is a way to assess 
the robustness of our result.
In the following, 
we will denote the fraction of superfluid density by
``superfluid fraction'' and we always refer to Leggett's 
upper bound to this quantity, unless otherwise specified.

The rest of this paper is organized as follows. In section~\ref{sec:bound}, we discuss how to adapt
Leggett's bound to an amorphous solid. In section~\ref{sec:Gaussian}, we compute the bound for 
a profile made of Gaussian fluctuations around a classical configuration, and compare the results
for an amorphous and an ordered solid, while in section~\ref{sec:numerical1} we discuss
previous numerical computations~\cite{BPS1}. 
In section~\ref{sec:numerical2} we try to obtain more precise information by comparing a classical simulation
of a glass-forming system with a PIMC numerical simulation of Helium. In section~\ref{sec:experimental},
we show that under some approximations one can obtain a formula for the bound that can 
-- at least in principle --
be computed from neutron or X-ray scattering data.

\section{Leggett's bound}
\label{sec:bound}

Leggett showed in his pioneering work on supersolidity that the wavefunction of the ground state of a system of bosonic 
particles inside a rotating cylindrical container can be obtained by finding the ground state for the non-rotating system but 
with new boundary conditions \cite{Le70}. Using cylindrical polar coordinates and assuming that the thickness of the cylinder 
is much smaller than the radius $R$, the new boundary conditions correspond to imposing
that the wave function gets an extra phase factor $\exp(-2\pi i m R^2 \omega/\hbar)$ when the angle $\theta_i$ of any particle $i$ is shifted by $2\pi$. 
Here $m$ is the particle mass and $\omega$ the radial velocity. 
From the $\omega$ dependence of the energy of the ground state, $E_{min}(\omega)$, 
obtained with these new boundary conditions one can compute the superfluid density $\rho_s$ by:
\[
\frac{\rho_s}{\rho}=\lim_{\omega \rightarrow 0} \frac{1}{I_0}\frac{\partial^2 E_{min}(\omega)}{\partial \omega^2}
\]
where $\rho$ is the particle density and $I_0 = N m R^2$ the classical moment of inertia. From this expression it is clear that 
upper bounds on the superfluid density can be obtained by using variational wavefunctions that in the 
$\omega\rightarrow 0$ limit tend to the wavefunction for a non-rotating container.  
Leggett used a variational wavefunction of the form 
$\Psi(\vec r_1,\cdots,\vec r_N)=\Psi_0(\vec r_1,\cdots,\vec r_N) \exp [ i \sum_i \f(\vec r_i)]$, where 
$\Psi_0$ is the ground state wavefunction for the non-rotating case and $\phi = \sum_i \f(\vec r_i)$ a sum of phases
satisfying the condition $\f(\theta)=\f(\theta+2\pi)-2\pi mR^2\omega/\hbar$~\cite{Le70,Sa76}. 
The bound can be improved by including two-body correlations \cite{SGR07}.
Defining 
\beq
\rho(\vec r) = \int d\vec r_1 \cdots d\vec r_N | \Psi_0(\vec r_1,\cdots,\vec r_N) |^2 
\sum_i \delta(\vec r-\vec r_i),
\eeq
which is the density profile in the ground state, one finds that the variational estimation of $E_{min}(\omega)$
reads:
\beq \label{var1}
E_{min}(\omega) = E_0+\frac{\hbar^2}{2m}\int d\vec r [\nabla \f(\vec r)]^2\rho(\vec r),
\eeq
where $E_0$ is the ground state energy in the non-rotating case.

Because of the assumption that the thickness of the cylinder is much smaller than the radius, 
one can simplify the problem even further by ``unrolling'' the annulus and consider the system
inside a parallelepiped of length $L=2\pi R$ in the $x$ direction. In this geometry the phase $\f$
has to satisfy the boundary condition $\f(0,y,z)=\f(L,y,z)-v_0L$ where $v_0=mR\omega/\hbar$.
The minimization of (\ref{var1}) with respect to $\f$ leads to the equation for
$\f(\vec r)$:
\beq\label{condi}
\begin{split}
\vec \nabla \cdot [\r(\vec r) \, \nabla \f(\vec r) ] = 0 \  \\
\end{split}\eeq
and results in an upper bound on the superfluid density:
\beq\label{boundr}
\r_s = \frac{1}{V v_0^2} \int_V d\vec r \r(\vec r) \, | \nabla \f(\vec r)|^2 \ .
\eeq
Note that if $\f_{v_0}(\vec r)$ is a solution of (\ref{condi}) with boundary conditions
$\f(0,y,z)=\f(L,y,z)-v_0L$, then $\f_{v_0'} = (v_0'/v_0) \f_{v_0}$ is a solution with boundary
conditions corresponding to $v_0'$. Hence, Eq.~(\ref{boundr}) does not depend on $v_0$
and we can choose $v_0 = 1$ without loss of generality. 
Furthermore, while in the geometry
described above the wavefunction should satisfy hard wall conditions at the boundary of
the box in the $y$ and $z$ directions, we will simplify the problem by considering periodic
boundary conditions in the $y$ and $z$ directions \cite{Forster}.

In order to find a solution of Eq.~(\ref{condi}) satisfying the correct boundary condition is useful
to rewrite $\f$ as 
\beq
\f(\vec r) = \vec v_0 \cdot \vec r + \d \f(\vec r), 
\eeq
where $\d\f(\vec r)$ is defined inside the volume $V$ and satisfies periodic boundary conditions,
and $\vec v_0$ is a unit vector. In the original problem $\vec v_0 = \hat x$, but since we reformulated
the problem in a periodic cubic box, the direction
of $\vec v_0$ can be varied without affecting the result, in the limit $V\to\io$.
Since $\d \f(\vec r)$ is periodic, we can write the equations in Fourier space (see Appendix~\ref{app:A} for details): 
\beq\label{condiF}
\vec q \cdot \vec v_0 \r_{\vec q} = \sum_{\vec p \neq \vec 0} (\vec q \cdot \vec p) \r_{\vec q-\vec p} \, i\d\f_{\vec p} \ ,
\eeq
and from the solution for $i\d\f_{\vec q}$ one can obtain the Leggett bound~\cite{Sa76}, that reads in Fourier space:
\beq\label{rs}
\frac{\r_s}{\r} = 1 - \frac1{\r v_0^2} \sum_{\vec q \neq \vec 0} (\vec v_0 \cdot \vec q) i \d\f_{\vec q} \r_{-\vec q} \ .
\eeq
Given the density profile, the linear equation (\ref{condiF}) for $i\d\f_{\vec q}$ can be solved by truncating the 
sum over momenta
at a given cutoff, $|\vec q| < q_{max}$, 
so that the problem reduces to solving a finite set of linear equations, which can be done by matrix inversion. 
We accomplish this via a LU decomposition~\cite{recipes}.

An important remark is that the truncation preserves the variational nature of the computation. Indeed, it can be
seen as setting $\d\f_{\vec q}=0$ for $|\vec q| \geq q_{max}$, which amounts to a particular choice of the variational
function $\d\f(\vec r)$ and hence still gives an upper bound on the true superfluid fraction.

Another important remark is that the bound derived above applies only, strictly speaking, to the true ground state
of the system. In the following however, we are interested in applying it to the glass state, which is at best
a long-lived metastable state, the crystal being always the true ground state. Still, it is clear from the derivation
that if the life time $\t$ of the state is very long, such that for any experimentally accessible frequency one
has $\omega \t \gg 1$, then the system does not have time
to escape from the metastable state during the experiment and the bound should apply without modification.

\section{Superfluid fraction of amorphous solids}

\subsection{Hard sphere systems}
\label{sec:Gaussian}

In order to understand whether disorder in the density profile can lead to an increase of the superfluid density,
we shall compare the result of the bound for an amorphous glassy profile and the corresponding crystal. 
The only input for our study are the density profiles of the amorphous and crystal state. Unfortunately,
the former is not available for He$^4$ in realistic conditions. As a consequence, we decided for a first study 
to focus on a more simple and academic case that can still provide insights on the role of disorder. 
We consider the amorphous and crystalline density profiles that one obtains for classical hard spheres. 
Although this certainly is not a realistic model of density profiles for He$^4$, it allows us to address the role
of disorder on $\rho_s$. Furthermore, a mapping from quantum systems at zero temperature and classical 
Brownian systems allows one to find quantum many particle models whose ground state wave-function can be mapped exactly on (the square of) the probability distribution of classical hard spheres systems~\cite{BCZ08}. Thus, 
the results of this section apply directly to those models.  

Classical hard spheres are known to be characterized by a high density
crystal FCC phase. However, if compressed fast enough, or due to a small polydispersity, the hard spheres
freeze in an amorphous glassy state. A typical density profile of a very quickly compressed glassy state can be 
obtained by the Lubachevski-Stillinger compression algorithm~\cite{LS90} (we used the implementation of \cite{DTS05c}), 
which is know to be very efficient in producing amorphous jammed configurations. The output of the algorithm 
are the positions $\RR = \{ R_1 , \cdots, R_N \}$ of the particles in a random close packed state (at infinite pressure).
The algorithm is deterministic, but different final configurations are obtained by starting the compression from random 
initial configurations of points.
The compression runs were performed at very fast rates 
(we fixed the parameter $\g=0.1$, see~\cite{DTS05c,PZ08} for details) 
in order to avoid crystallization. 

\begin{figure}\label{cryst}
\includegraphics[width=.5\textwidth]{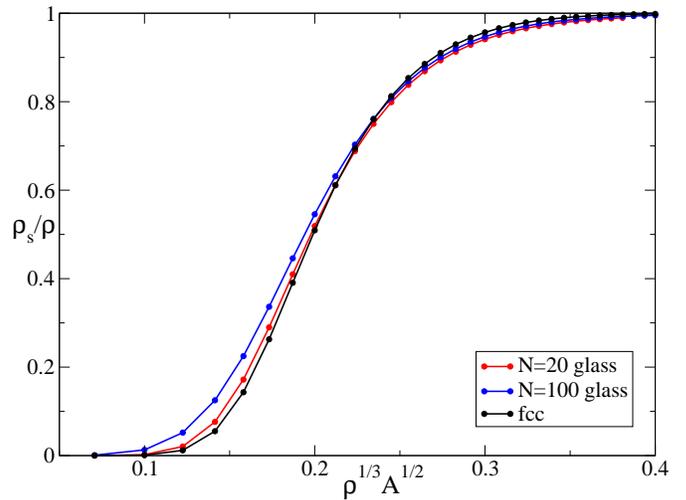}
\caption{
Leggett upper bound for $\r_s/\r$,
for a Gaussian profile of width $A^{1/2}$ around an amorphous jammed configuration and in a FCC lattice,
as a function of the adimensional parameter $\ell = \r^{1/3} A^{1/2}$ (the Lindemann ratio).
}
\label{fig:1}
\end{figure}

Furthermore, we will assume that the density profile of a typical glassy configuration at finite pressure is 
the sum of Gaussians centered around the amorphous sites, which are the output of the previous algorithm.
For classical systems, this assumption has been tested numerically for FCC crystals \cite{YA74},
and has been often used in density functional computations
of both ordered~\cite{DAC95} and amorphous structures~\cite{SW84,PZ08}, giving accurate results.
For quantum systems, 
the Gaussian model has been shown to be accurate enough, 
at least for the purpose of computing the Leggett's upper bound~\cite{FP74,SJ06,GRS07}.

\begin{table*}
\begin{tabular}{|c|cc|c|c|}
\hline
 & HCP, Leggett's bound & (Ref.\cite{GRS07}) & Glass, Leggett's bound (this work) & Glass, QMC (Ref.~\cite{BPS1})  \\
$\r$ ($\AA$$^{-3}$) & $\ell$ & $\r_s/\r$ & $\r_s/\r$ & $\r_s/\r$ \\
\hline
0.029 & 0.167 & 0.22 & 0.282 & 0.6 \\
0.0353 & 0.143 & 0.06 & 0.127 & 0.07 \\
\hline
\end{tabular}
\caption{
Leggett's bound for He$^4$ in the HCP crystal state~\cite{GRS07} and glassy state. Quantum Monte Carlo results
for the glass are also reported~\cite{BPS1}.
}
\label{tab:1}
\end{table*}

For a given configuration $\RR$, the density profile we use is defined as
\beq
\rho(\vec r | \RR) = \sum_i \g_A(|\vec r - \vec R_i|) = \int_V d\vec s \, \g_A(|\vec r - \vec s|) \sum_i \delta(\vec s - \vec R_i) \ ,
\eeq
where $\g_A(\vec x) = \exp( -|\vec x|^2/ (2 A) )/(2 \pi A)^{3/2}$ 
is a normalized Gaussian of width $A$, and $|\vec r - \vec R_i|$ is the distance on the periodic box, \ie it
is the distance between $\vec r$ and its closest image of $\vec R_i$.
The corresponding Fourier transform reads (neglecting terms of order $\exp(-L^2/A)$):
\beq\label{rho}
\rho_{\vec q}(\RR) = e^{-A q^2/2} \frac1V \sum_i e^{i \vec q \cdot \vec R_i} \ .
\eeq

In solving Eqs.~(\ref{condiF}) and (\ref{rs}) we considered amorphous
configurations of $N=20$ and $N=100$ particles. 
All the calculations were done with the cut-off set at $q_{max} = 20\pi/L$.
We checked that the result does not depend on the specific amorphous configuration used by considering
different amorphous configurations $\RR^\a$, $\a = 1, \cdots, \NN$; this is expected since the superfluid
density is a macroscopic quantity. The reported results are therefore averaged over $10$ independent 
configurations. More details on the numerics can be found in Appendix~\ref{app:A}.

The results are plotted in Figure~\ref{fig:1}. One can notice that, apart from the smallest values of the 
dimensionless parameter, the two curves corresponding to 20 and 100 particle configurations perfectly
agree. The discrepancy in the region of small $\ell = \rho^{1/3}A^{1/2}$ is due to the approximation
brought by the introduction of a cut-off, and vanishes in the limit $q_{max} \gg 1/\sqrt{A}$.

In order to understand to what extent the disorder influences the value of the superfluid density, 
we compare the superfluid fraction found in the amorphous
system to the values obtained through the same calculations in the case of a 
crystal~\cite{Sa76,FP74,SJ06,GRS07}.
Figure~\ref{fig:1} reports the results for the average superfluid fraction of the amorphous solid
just described and those
corresponding to the FCC lattice (which is the thermodynamically stable one for hard spheres) 
for the $\vec R_i$, according to the same Gaussian model (in the latter case our results are consistent
with previous ones~\cite{Sa76,FP74,SJ06,GRS07}).
The difference between the two is very small, suggesting two conclusions.
\begin{enumerate}
\item Disorder does not 
influence much the superfluid behavior of the system for comparable values of $ \rho^{1/3}A^{1/2}$, 
at least at the level of this variational calculation.
\item The dependence of $\rho_s$ 
on the density profile is mainly through the Lindemann ratio $\ell = \rho^{1/3}A^{1/2}$. 
This conjecture allows us to obtain an estimate of the Leggett upper bound for $\rho_s$ 
in more realistic cases as we will do in the next section. 
\end{enumerate}
To conclude this section, we observe that the above results allow to obtain a quantitative upper bound
for the superfluid fraction of a system whose wavefunction is exactly the Jastrow wavefunction
corresponding to classical hard spheres. The quantum glassy phase of this system has been discussed 
in~\cite{BCZ08}. In both the crystal and glassy phases, the values of $A^{1/2}$ for classical hard spheres
do not exceed $0.1$ (in units of the sphere diameter)~\cite{YA74,DAC95,PZ08}, 
and the same is true for $\ell$, since the density is very close to 1 (in the same units) in both solid
phases. Using the results of Fig.~\ref{fig:1}, we obtain an upper bound $\r_s/\r \lesssim 0.1 \%$, which is
consistent with the extremely small values of the condensate fraction found in~\cite{BCZ08}.

\subsection{Superfluid fraction of amorphous solid Helium 4}
\label{sec:numerical1}

In this section, we attempt an application of our results to the more interesting case of
disordered solid He$^4$, based on the observation above, that an estimate of the Lindemann
ratio $\ell = \r^{1/3} A^{1/2}$, together with the results of Fig.~\ref{fig:1}, should provide
a reasonable estimate of Leggett's bound.

At the end of Ref.\cite{GRS07} it is stated that, by fitting the Path Integral Monte Carlo density profile
of HCP solid He$^4$, one obtains a value $\sqrt{A} = 0.1274 \, d$ at $\r = 0.0353$~$\AA$$^{-3}$ and
$\sqrt{A} = 0.1486 \, d$ at $\r = 0.029$~$\AA$$^{-3}$. Here $d$ is the nearest-neighbor distance for the
HCP lattice. The number density of the HCP lattice satisfies the relation $\r d^3 = \sqrt{2}$, hence
$d = 2^{1/6}/\r^{1/3}$ and $\ell = \sqrt{A} \r^{1/3} = 2^{1/6} \sqrt{A}/d$.
In the same reference
it is also stated that the upper bound computed by using the fitted Gaussian density profile coincides
with the one obtained by using the true PIMC density profile, and corresponds respectively 
to $\r_s/\r = 0.06$ and $0.22$. These values are reported in table~\ref{tab:1}.

We now make the following assumptions:
\begin{enumerate}
\item {\it At least for the purpose of computing Leggett's upper bound},
the true density profile can be fitted to a Gaussian profile. This is true for the crystal~\cite{GRS07}
and we assume that it remains true for an amorphous solid.
\item The parameter $\ell$ for the amorphous solid is smaller than that of the crystal at the same density.
This can be understood by observing that crystalline
configurations are better packed than amorphous configurations, therefore leaving more room (``free volume'')
for fluctuations. It is true for Jastrow wavefunctions~\cite{BCZ08} (\ie classical system) and we do not find
any reason why quantum fluctuations should dramatically affect this property.
\end{enumerate}
Based on these assumptions, the true Leggett's bound for the amorphous system should be smaller than the same
bound for the crystal at the same density. This can be estimated using the values of $\ell$ reported in~\cite{GRS07}
and reading the corresponding superfluid fraction from Fig.~\ref{fig:1} or using the results obtained in~\cite{GRS07}
for the HCP crystal. These values are reported in table~\ref{tab:1} and are similar. 

We compare the upper bound obtained in this way with the values of $\r_s$ obtained numerically by 
Boninsegni et al. via PIMC~\cite{BPS1}. 
Interestingly, we find that the bound is very close to the PIMC numerical result, 
and in particular at the smallest density the bound is violated by the PIMC result. 
This can be due either to the very rough approximations involved in our computation,
or to the fact that the glass is not a really long-lived metastable state at this very low density.
The latter possibility, i.e. that the system is rapidly evolving out of equilibrium,
would invalidate the derivation of Leggett's bound but it would also raise problematic questions 
regarding the measurement of $\rho_s$ using the Ceperley formula,  which is strictly valid 
if thermodynamic equilibrium is achieved and in the limit of small frequency.

\section{Does a stable glass state exists for Helium 4?}
\label{sec:numerical2}

\begin{figure}
\includegraphics[width=.475\textwidth]{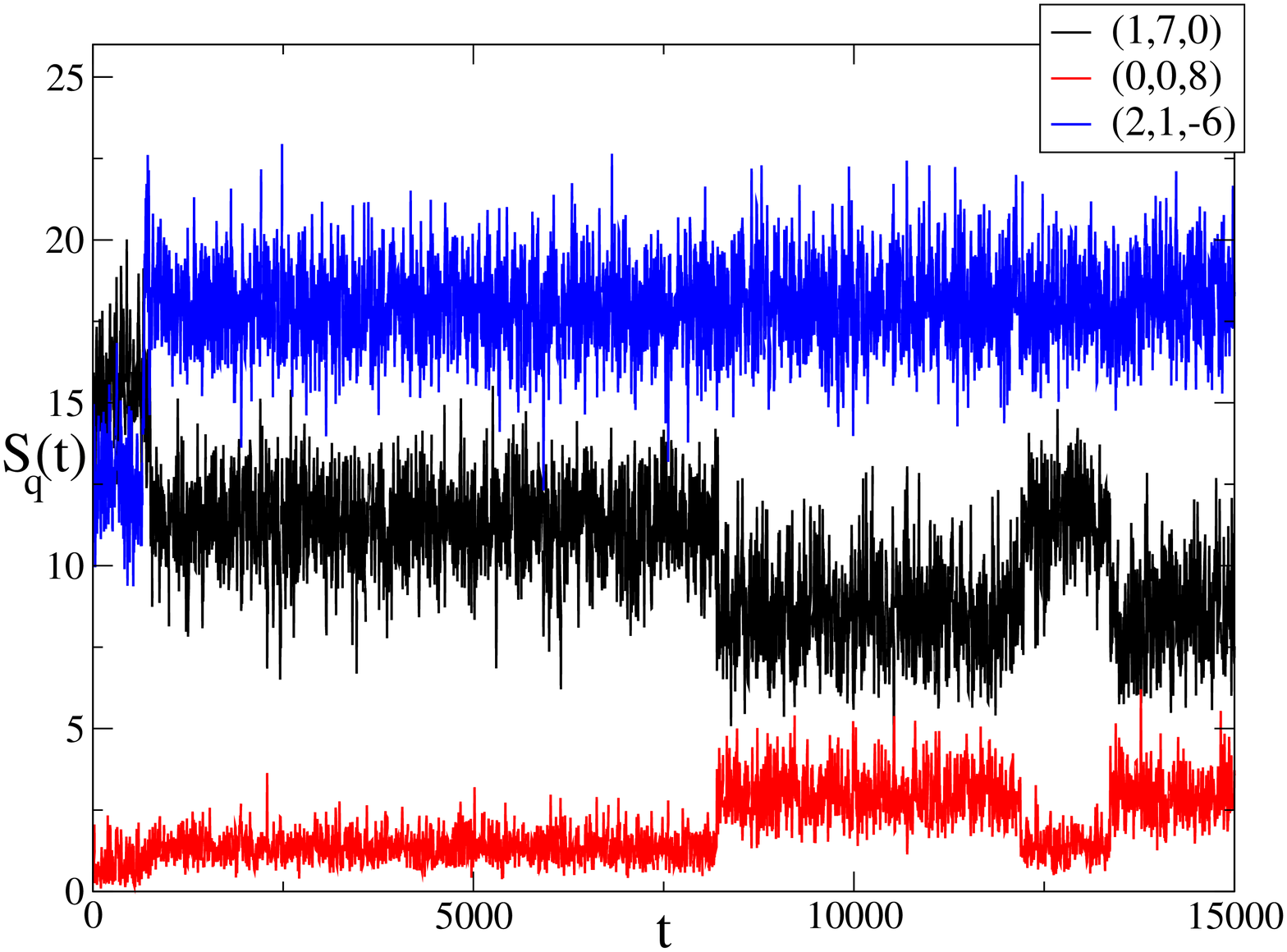}
\includegraphics[width=.475\textwidth]{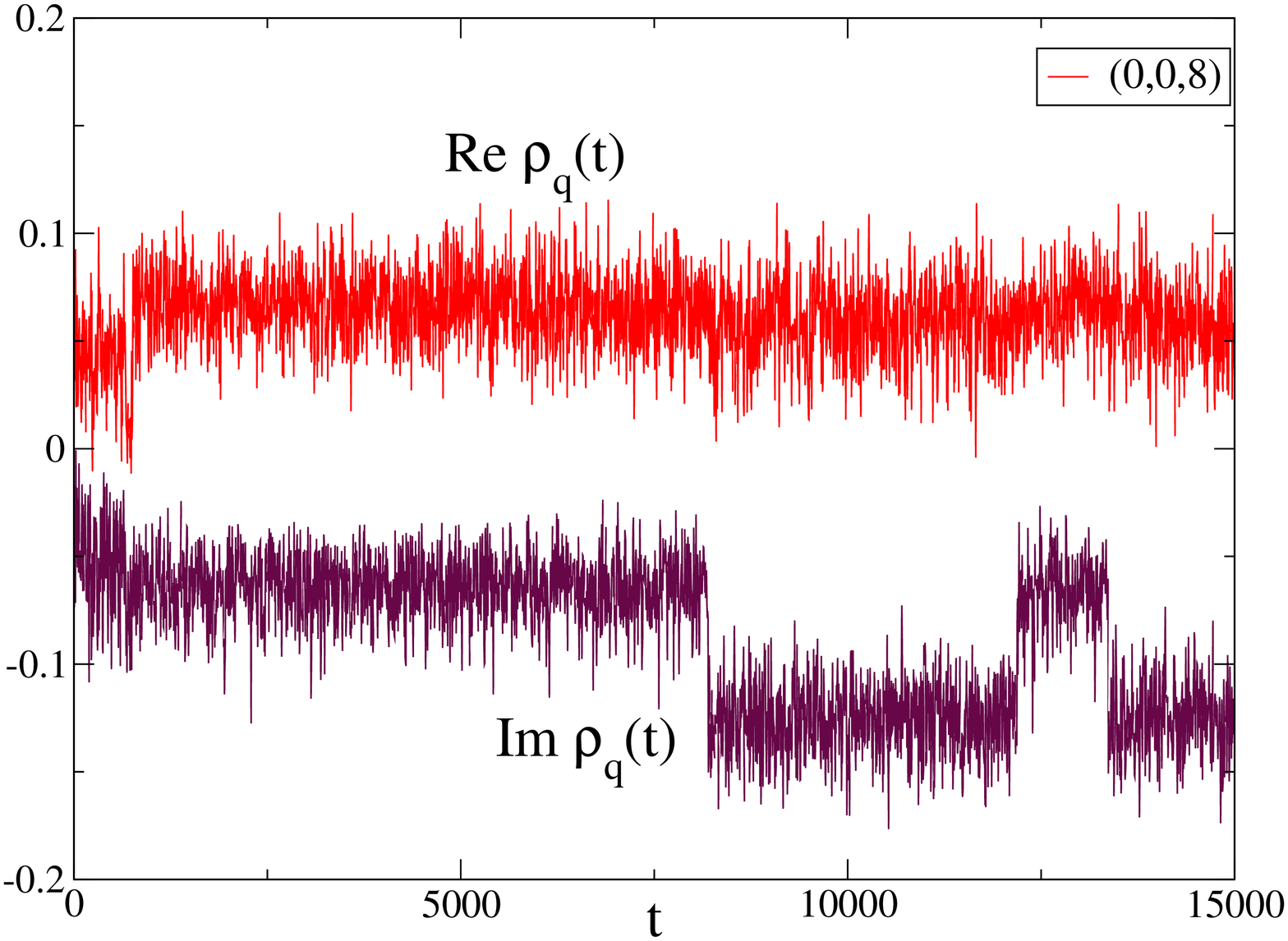}
\includegraphics[width=.475\textwidth]{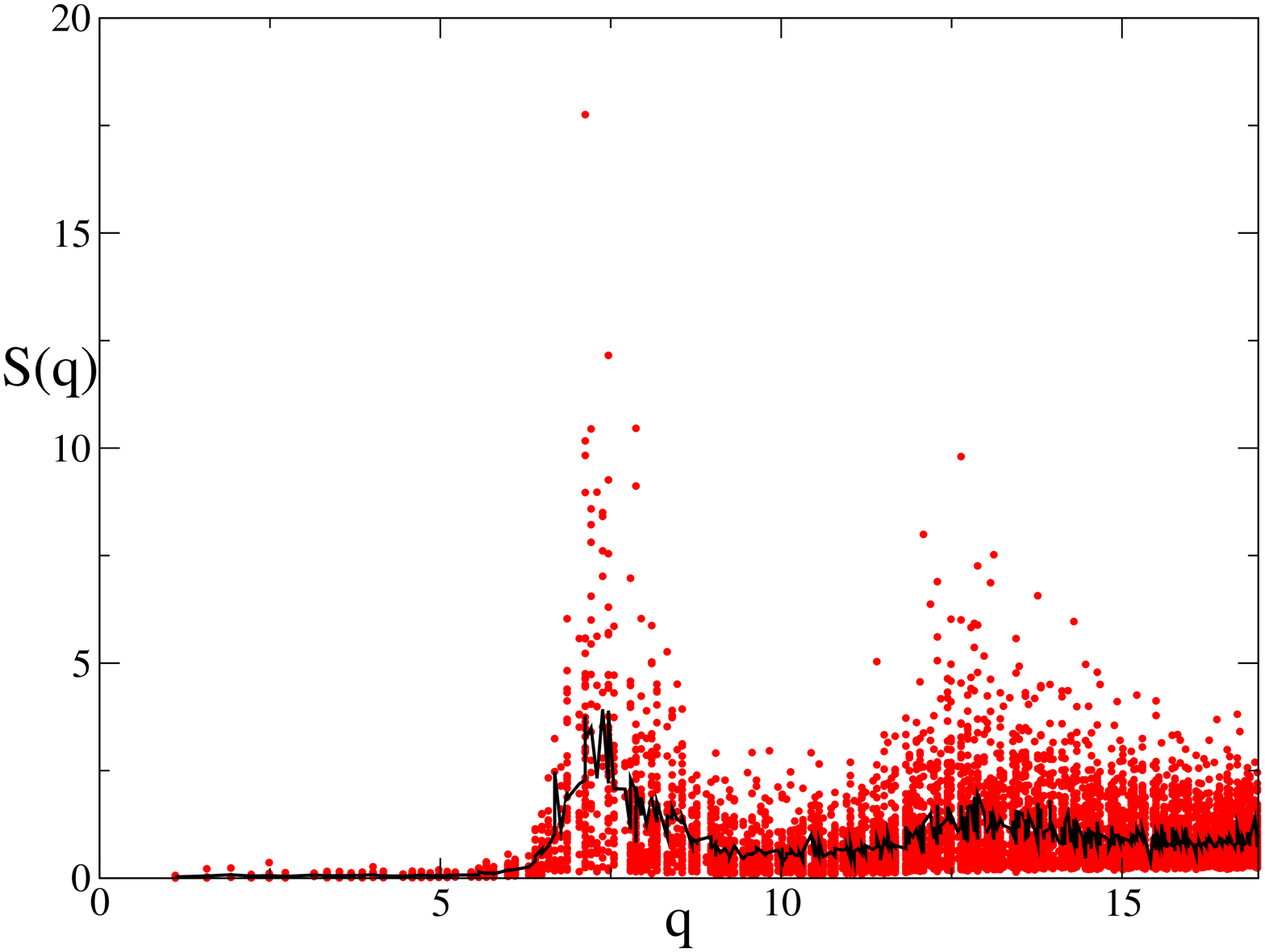}
\caption{
Evolution of the density profile after a quench from high
to low temperature for a classical glass forming system,
using Molecular Dynamics.
(Top) Instantaneous value of $S_{\vec q}(t)$ for three
representative 
values of $\vec q$ (the corresponding $(n_x,n_y,n_z)$ are indicated
in the caption). 
(Middle) Instantaneous values of $\r_{\vec q}(t)$ for a representative
value of $\vec q$.
(Bottom) The time average of $S_{\vec q}(t)$ over
the whole simulation, as a function of $q$ (in reduced LJ units).
Scatter points are values for a given $\vec q$, the full black line is the angular
average over all vectors with the same modulus.
}\label{Sq_classical_glass}
\end{figure}

In order to study the stability of the glass phase in Helium 4,
we performed Path Integral Monte Carlo simulations, that we discuss
in this section. Before discussing the more complex quantum simulation,
we present some classical simulations in order
to deal with a well controlled situation, where the presence
of a glass transition has been firmly established.

\subsubsection{What should we expect from a glass-forming system? A classical simulation}

We performed standard Molecular Dynamics (MD) simulations
of the Kob-Andersen binary mixture~\cite{KA95},
which is known to be a good glass former and does not show
any sign of crystallization even after very long MD runs at
low temperature. The latter is a mixture of two types of particles
(A and B), interacting through different Lennard-Jones potentials, 
with the parameters specified in~\cite{KA95}. 
In the rest of this section we use reduced
Lennard-Jones units, namely we use $\s_{AA}$ and $\e_{AA}$ as
units of length and energy, and $m$ as unit of mass. Consequently,
$\sqrt{m \s_{AA}^2/\e_{AA}}$ is the unit of time (the latter 
convention is slightly different from the one of~\cite{KA95}).
Note that to compare with Helium one should keep in mind that
for that system $\s \sim 2.56 \, \AA$ and $\e \sim 10.2$ K.

We quenched a dense ($\rho=1.2$) system of $N=216$ particles 
from very high temperature
($T=2$) to very low temperature ($T=0.05$) deep in the glass phase (the glass
transition temperature being around $T = 0.435$ at this density~\cite{KA95}).
We run the simulation for a total time $\t = 15000$ and
we printed configurations every $\D t=5$ which is of the order of the
decorrelation time in the glass (estimated from the decay of the
self scattering functions). From each configuration we deduced
\beq
\r_{\vec q}(t) = \frac{1}{V} \sum_j e^{i \vec q \cdot \vec r_j(t)} \ ,
\eeq
where $\vec r_j(t)$ is the position of particle $j$ at time $t$,
and the corresponding instantaneous value of the static 
structure factor $S_{\vec q}(t) = V |\r_{\vec q}(t)|^2/\r$. 

In Fig.~\ref{Sq_classical_glass} 
we plotted $\r_{\vec q}(t)$ and the structure factor $S_{\vec q}(t)$
as a function of MD time after the quench. 
The vectors $\vec q = 2\pi/L (n_x,n_y,n_z)$ and the corresponding integers
are given in the caption. 
We see that after a short transient, the density
profiles fluctuate around a non-zero value which is quite stable, except
for some rare ``crack'' events where the density changes abruptly. 
These are probably due to groups of particles that switch back and
forth between two different locally stable configurations.
This system is indeed extremely dense and at very low $T$, therefore its dynamics is 
basically that of harmonic vibrations around local minima of the potential 
(except for the rare cracks).
The largest instantaneous value of $S_{\vec q}(t)$ corresponds
to the $(2,1,-6)$ curve in Fig.~\ref{Sq_classical_glass} for all $t>1000$;
therefore, all values are smaller than $20$ at all times, showing that
there are no Bragg peaks.
This is what we expect to see in a glass. In this case, we can easily deduce
the average values of $\rho_{\vec q}$ for a given glassy configurations by
taking the average of $\r_{\vec q}(t)$ over a time interval where there are 
no crack events. From these, we could compute the Leggett bound as previously
discussed.

\subsubsection{Absence of a stable glass phase from a Path Integral Monte Carlo simulation}

\begin{figure}
\includegraphics[width=.475\textwidth]{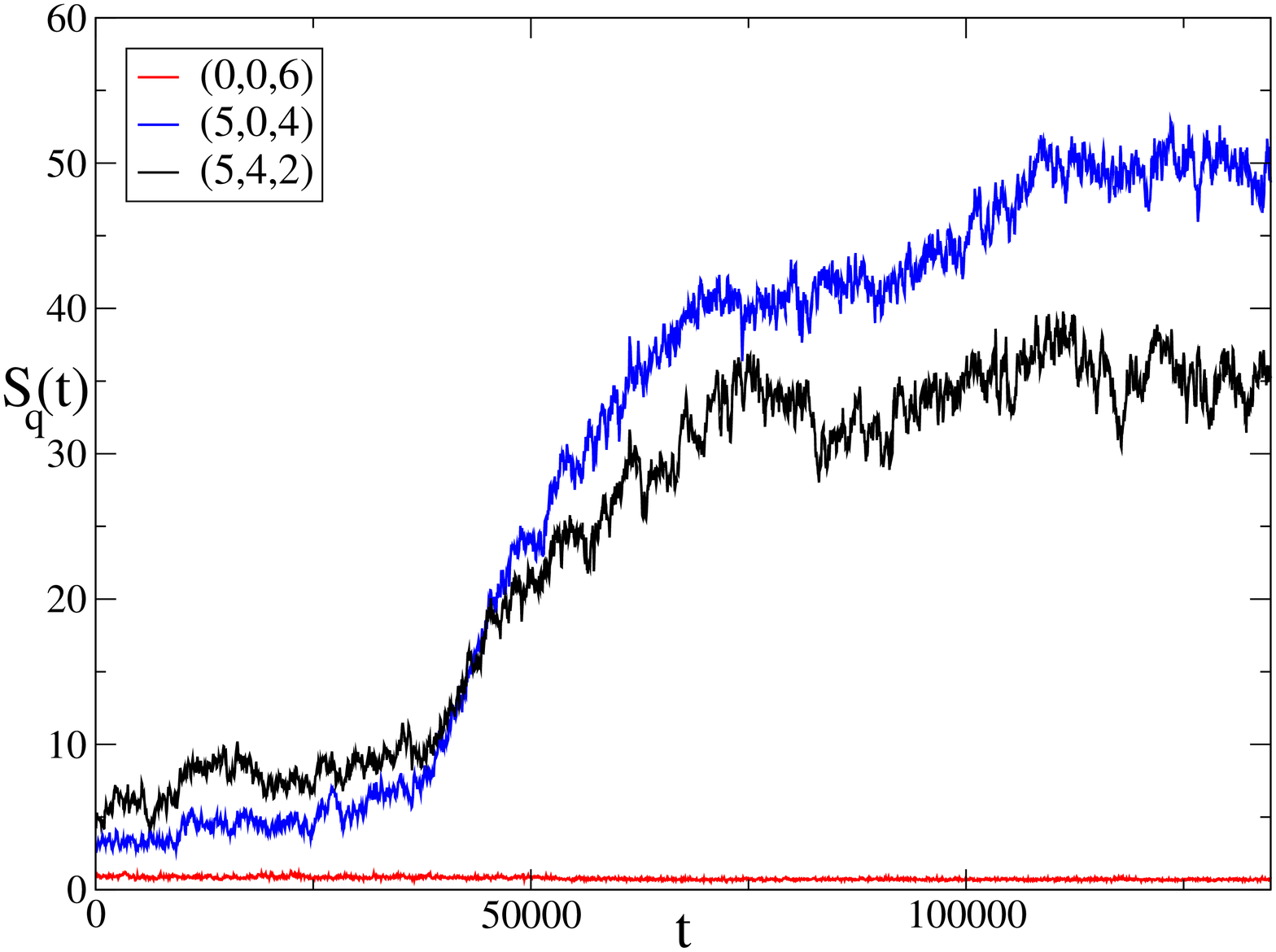}
\includegraphics[width=.475\textwidth]{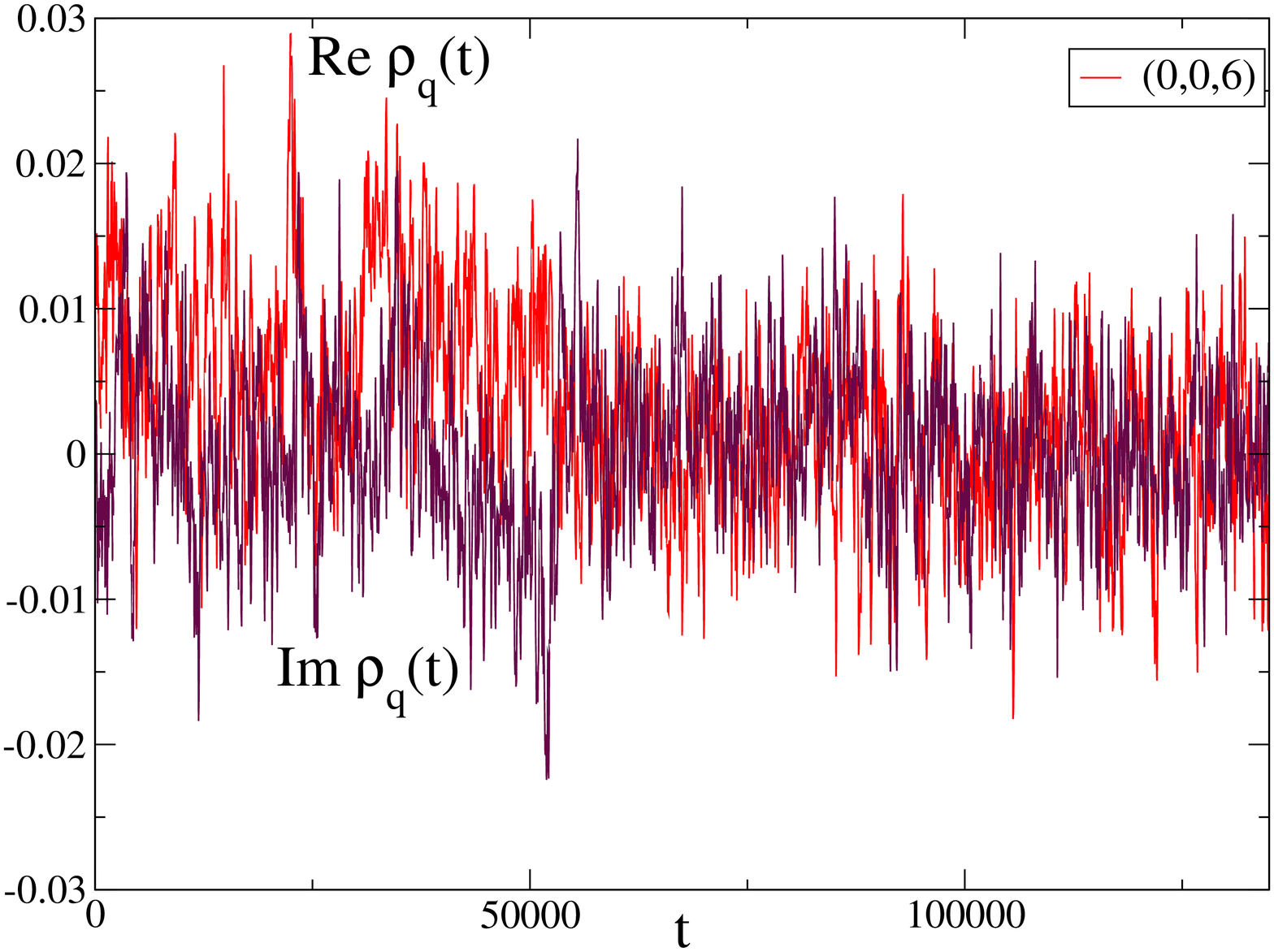}
\includegraphics[width=.475\textwidth]{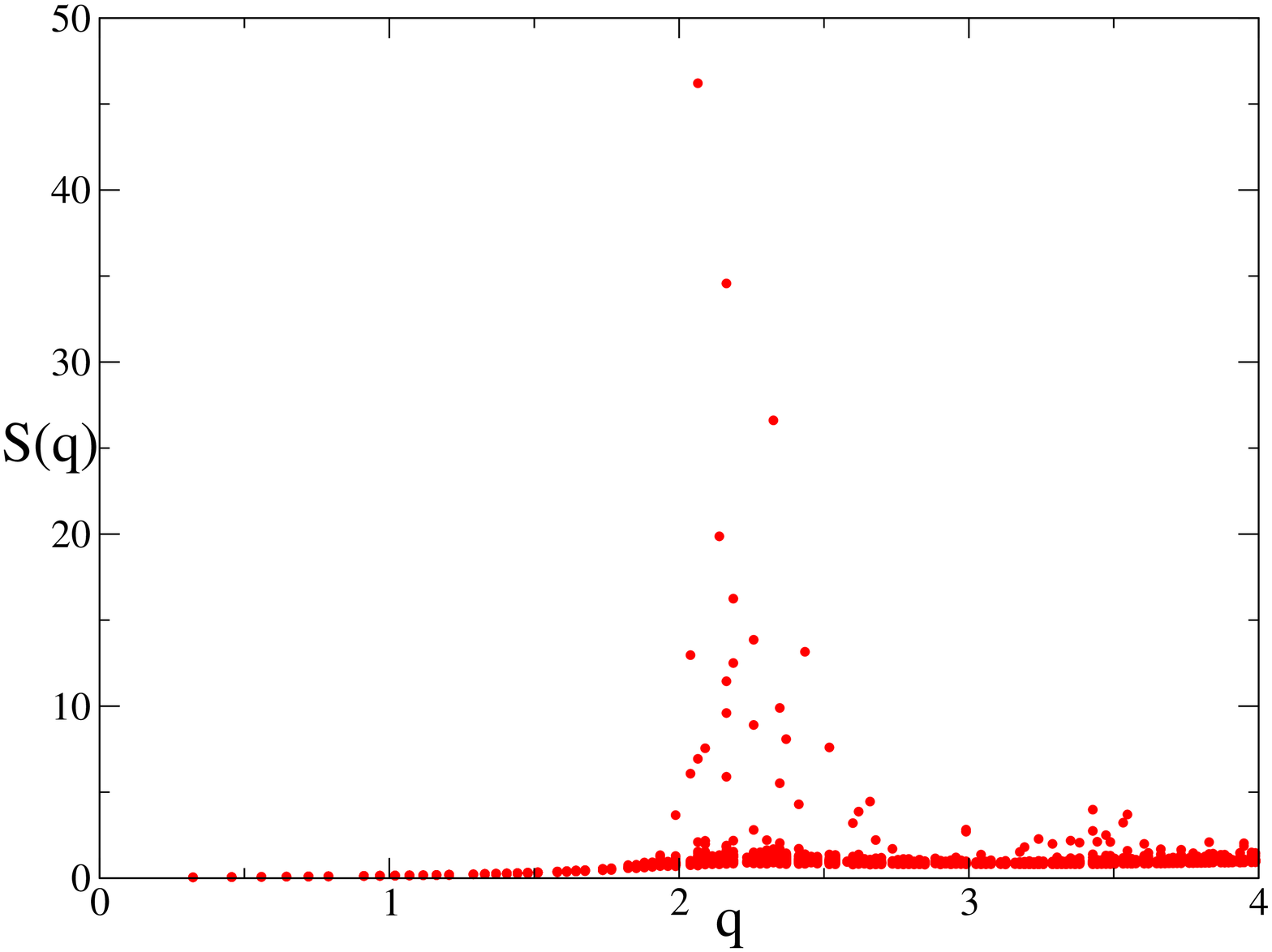}
\caption{
Evolution of the density profile after a quench from high
to low temperature for a quantum Helium 4 system,
using Path Integral Monte Carlo. Time here represents
the number of Monte Carlo sweeps.
The panels are the same as in Fig.~\ref{Sq_classical_glass},
except that the average of $S_{\vec q}(t)$ in the lower panel
has been taken for $t > 75000$, and the angular average is
not reported because of the strong anisotropy of the result.
All quantities are plotted using $\AA$ as units of length.
}
\label{Sq_quantum_glass}
\end{figure}

Motivated by results of~\cite{BPS1} we tried to compute
the superfluid fraction based directly on Path Integral Monte Carlo data.
Unfortunately, PIMC does not give access to the
real time dynamics of the system, but following~\cite{BPS1} we studied the
Monte Carlo dynamics, in the hope that this is a reasonable proxy
for the real time dynamics.

The representation of quantum systems in PIMC involves certain
important extensions
beyond the classical representation of point particles.
To begin with, particles are
represented by paths (or polymers) in space.  These paths manifest
the zero point motion inherent in the quantum mechanical system.
For distinguishable particles, this is the only difference. For
particles with statistics (bosons), these paths then can permute
onto each other forming larger paths or cycles.

We initially focus on studying a quenched quantum system of Helium particles
but require that they act like distinguishable particles. There are a number
of potential advantages of this approach. To begin with, one may hope
that distinguishable particles are more likely to retain the relationship
between real dynamics and the Monte Carlo dynamics. Secondly, the
simulation of distinguishable particles is faster and more easily
parallelized over many processors allowing for longer simulations.

We used the Aziz potential as a model for Helium~\cite{Ce95}, and in
this section we always use Angstroms as units of length and Kelvins as units
of temperature.
The pair product action is used as the approximation for the high temperature
density matrix and an imaginary time step of 
$\delta \tau=0.025$ K is used.  We equilibrated
a system of $N=216$ particles in the liquid phase 
at a density of 0.029 $\AA$$^{-1}$ and a temperature of $T=2$ K.
The system is then instantaneously quenched to $T=0.166$ K.  This is
accomplished by taking a snapshot of the paths from $T=2$ K and then,
for each time slice of the old path, placing 12 time slices for the
new lower temperature path; this is similar to what was done by 
Boninsegni et al. \cite{BPS1}.  
We then run the PIMC from this quenched
configuration.  These paths are obviously highly artificial
because the distances between many adjacent time slices are zero.
Over a very short period at the beginning of the quenched run, though,
this artificial aspect of the path quickly relaxes leaving the paths in a
configuration that mirrors the higher temperature formation.

In the following we refer to $t$ as the PIMC ``time'' 
(number of PIMC sweeps\footnote{We define a sweep as attempting 
a displace move on (an expected) $10\%$ of the particles 
and attempting bisection moves on (not necessarily unique) $0.4 N/(T \, \d\tau)$ time slices.}), 
while $\t$ is the imaginary time. At each ``time'' $t$, the PIMC code returns
a configuration $\vec r^\t_j(t)$, the latter being the imaginary time trajectory
of particle $j$ as function of the imaginary time $\t$. 
We can define the instantaneous density as
\beq
\r_{\vec q}(t) = \frac{1}{\b V} \sum_j \int_0^\b d\t \, e^{i \vec q \cdot \vec r^\t_j(t)} \ ,
\eeq
and the instantaneous structure factor
\beq
S_{\vec q}(t) = \frac{1}{\b N} \sum_{j,k} \int_0^\b d\t \, e^{i \vec q \cdot [ \vec r^\t_j(t) -\vec r^\t_k(t) ] } \ .
\eeq
Note that in the quantum case, at variance with the classical case, these two quantities are not directly related.
At each PIMC sweep we recorded the values of the above quantities, which we then averaged over 50 PIMC
sweeps in order to eliminate part of the fluctuations. 

The results for a representative run of the above procedure 
are reported in Fig.~\ref{Sq_quantum_glass}.
Unfortunately, the dynamics of this system looks quite different from the formation of a
glass from a quenched liquid.
First of all, the structure factor becomes
quite large for some values of $\vec q$, therefore suggesting the presence of large crystallites in the 
sample. Indeed, the largest value of the structure factor corresponds to the $(5,0,4)$ curve in Fig.~\ref{Sq_quantum_glass} 
at large times
and to the $(5,4,2)$ curve in Fig.~\ref{Sq_quantum_glass} at short times. We see that while at short times
the values of $S_{\vec q}(t)$ are smaller than 10, at larger times they grow up to 50, which clearly indicates the presence of large
crystallites in the sample (note in addition that these values have been averaged over 50 PIMC sweeps and also
over imaginary time).
Moreover, the $\rho_{\vec q}(t)$ (reported for a representative value of $\vec q$ in the middle panel of Fig.~\ref{Sq_quantum_glass}) 
are not fluctuating around some stable value; they display a sluggish 
evolution that does not allow us to identify a region of times where the system is close to some 
metastable density profile that does not evolve in time. What we can learn from this is that
the quenching from a (exchange-free) liquid to a (exchange-free) low temperature liquid froze to 
a (possibly very broken) crystal relatively quickly without showing any intermediate signs of
glassiness. Note however that this behavior was not observed in all runs: some runs did not display
signs of crystallization for times up to $\sim 200000$ PIMC sweeps. Still the dynamics was sluggish enough
to prevent the identification of a stable glass phase.
We also tried turning off some moves (the displace moves) in order to slow down the 
relaxation to the crystal, but the system still seems to freeze just as quickly.

In conclusions, we were not able to find a long-lived metastable glassy state in our quantum 
simulations. This is probably due to the fact that monodisperse systems always crystallize quite fast.  
This is well known in the classical case and seems to also hold true when quantum zero point
motion is introduced (at least in this specific example).
This leaves the discrepancy between our findings and those of~\cite{BPS1} to be explained.
One possibility is that exchange, that we neglected, may be critically important
for exhibiting the glassy behavior of Helium 4: it could be that the path integral 
at the low temperatures we are focusing on 
is dominated by exchange paths, whereas the paths that make the glass unstable are 
mainly without exchange; indeed we find them with our PIMC. In this case, the instability of the glass
would be a much rarer process once one takes into account exchange paths. In particular, since
crystals have a very low or zero superfluid fraction, we know that their corresponding path integral 
is dominated by paths without exchange. In consequence, eliminating the exchange could also make 
crystal nucleation easier since it makes it a less rare process. 


An additional possibility is that the glassy behavior is sensitive to the specific
details of the simulation (type of Monte Carlo moves, length of the paths, etc.).
We leave a more detailed investigation of this point for future study.

\section{Towards a method for experimentally assessing the Leggett bound}
\label{sec:experimental}

As we discussed previously, the problem in applying our analysis to realistic system is that the amorphous 
density profile of He$^4$ cannot be easily measured experimentally. Below, we endeavor to connect the bound on $\rho_s$ 
to the so-called non-ergodic factor $\wt g_q$, which in principle could be measured in experiments, 
\eg by neutrons or X-ray scattering. 
It is defined as
\beq
\frac{\r^2}{N} \wt g_q = \frac1{\cal N} \sum_\a \r_{\vec q}^\a \r_{-\vec q}^\a = \overline{\r_{\vec q} \r_{-\vec q}} \ ,
\eeq
where the overbar denotes the statistical average over the amorphous states 
sampled statistically by the system. These are indexed by $\a = 1,\cdots,\NN$, 
and under the Gaussian approximation each profile $\r_{\vec q}^\a$ is obtained from 
Eq.~(\ref{rho}) by plugging the reference positions corresponding to each different
amorphous configuration $\RR^\a$. The statistical average is performed with the weights $\alpha$ 
that correspond to the frequency with which they appear in an experiment, or equivalently their Boltzmann weight. 

First, let us focus on $\overline \r_s$, which is the average of the superfluid density
$\rho_s^\alpha$ corresponding to each amorphous state.
Since the superfluid density is a macroscopic quantity we expect (and we have checked
numerically, see Appendix~\ref{app:A}) a self-averaging behavior, \ie the fluctuations of $\rho_s^\alpha$ are negligible. 
However, as usual for disordered systems, the computations are easier for $\overline \r_s$.
Multiplying Eq.~(\ref{condiF}) by $\r_{-\vec q}^\a$ and averaging over $\a$ we obtain
\beq\label{condiFqp}
(\vec q \cdot \vec v_0 ) \frac{\r^2}{N} \wt g_q  = \sum_{\vec p \neq \vec 0} (\vec q \cdot \vec p) F(\vec q,\vec p) \ ,
\eeq
where we define, for $\vec p , \vec q \neq 0$ (that are the only cases involved in the equation above)
\beq
 F(\vec q,\vec p) = \frac1{\cal N} \sum_\a \r^\a_{\vec q-\vec p} \, i\d\f^\a_{\vec p} \r_{-\vec q}^\a=
 \overline{ \r_{\vec q-\vec p} \, i\d\f_{\vec p} \r_{-\vec q} } \ .
\eeq
Clearly $i \f_{\vec q}$ is strongly correlated to $\r_{\vec q}$, being the solution of (\ref{condiF}). In order to simplify
the problem we assume that these variable are Gaussian distributed.
Using Wick's theorem, one has
\beq\begin{split}
 F(\vec q,\vec p) & = 
 \overline{ \r_{\vec q-\vec p} } \hskip5pt \overline{ i\d\f_{\vec p} } \hskip5pt \overline{ \r_{-\vec q} } +
 \overline{ \r_{\vec q-\vec p} } \hskip5pt \overline{ i\d\f_{\vec p} \r_{-\vec q} } \\ &+
 \overline{ \r_{\vec q-\vec p}  i\d\f_{\vec p}} \hskip5pt \overline{ \r_{-\vec q} }+
 \overline{ \r_{\vec q-\vec p}  \r_{-\vec q}} \hskip5pt \overline{ i\d\f_{\vec p} } \ .
\end{split}\eeq
Note that, due to translation invariance of the averages over $\a$, one has
$\overline{ \r_{\vec q} } = \r \delta_{\vec q,\vec 0}$ and $\overline{ \r_{\vec q}  \r_{-\vec p}} = \frac{\r^2}{N} \wt g_q \delta_{\vec q,\vec p}$.
Hence, for $\vec p , \vec q \neq 0$, we get
\beq\label{gaussian}
 F(\vec q,\vec p) = 
 \overline{ \r_{\vec q-\vec p} } \hskip5pt \overline{ i\d\f_{\vec p} \r_{-\vec q} } = \r \delta_{\vec q,\vec p} \,
 \overline{ i\d\f_{\vec q} \r_{-\vec q} } \equiv  \r \delta_{\vec q,\vec p} \, F(\vec q)
\ .
\eeq
Substituting the last expression in (\ref{condiFqp}), we obtain
\beq
F(\vec q) = \frac{\r (\vec q \cdot \vec v_0 )\wt g_q}{N q^2} \ .
\eeq
Averaging (\ref{rs}) over $\a$, we get
\beq\label{verysimple}
\frac{\overline{ \rho_s}}{\r} = 1 - \frac1{\r v_0^2} \sum_{\vec q \neq \vec 0} (\vec v_0 \cdot \vec q) F(\vec q)
= 1 - \frac1N \sum_{\vec q\neq \vec 0} \frac{(\vec v_0 \cdot \vec q)^2}{v_0^2 q^2} \wt g_q \ .
\eeq
In the thermodynamic limit, the sum can be replaced by an integral, and performing the angular integration
we obtain:
\beq\label{boundapprox}
\frac{\overline{ \rho_s}}{\r} = 1 -\frac23 \int_0^\io \frac{dq \, q^2}{(2 \pi)^2 \r} \wt g_q \ .
\eeq
The same result can be obtained by means of a large $A$ expansion of the system of equations, which however
is poorly convergent and cannot be used in a systematic way, see Appendix~\ref{app:B}.

\begin{figure}
\includegraphics[width=.5\textwidth]{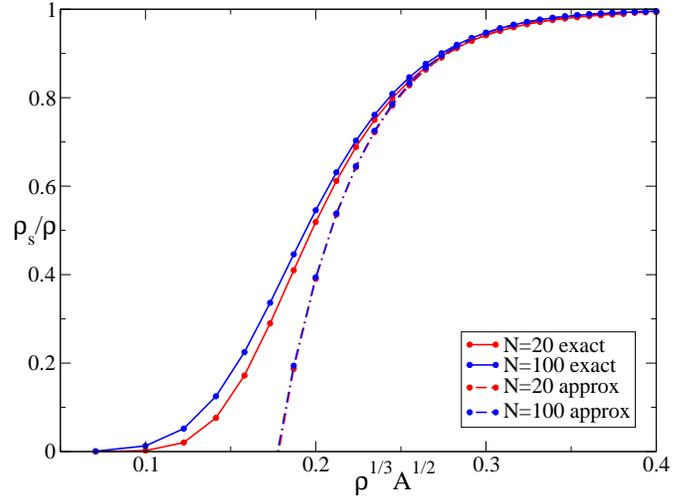}
\caption{
Result for $\r_s/\r$ as a function of $\ell = 
\r^{1/3} A^{1/2}$,
where $\vec R_i$ are the center of the spheres in an amorphous jammed configuration of $N$ spheres
with periodic boundary conditions. We report the exact computation according to Eq.~(\ref{condiF})
and the approximate result Eq.~(\ref{boundapprox}).
}
\label{fig:2}
\end{figure}

As before we need to introduce a cut-off in the sum on $\vec q$ in (\ref{rho}) and calculate numerically 
the non-ergodic factor $\wt g_q$ by averaging the density 
over the same configurations $\RR^\a$ considered above.
We set the cutoff according to the spherical constraint $|\vec q| \leq q_{max}$ . We increased $q_{max}$ until  
$q_{max}=20\pi/L$, when the convergence in $\wt g_q$ was reached. 
For the purpose of computing the non-ergodic factor and then the approximate bound, as given in Eq.~(\ref{verysimple}), 
 we averaged over 100 different configurations. In this case, in fact, one  does not face the computational problem of 
 inverting the linear system (\ref{condiF}) and thus a larger statistics can easily be taken.
 The results of the computations are shown in Figure \ref{fig:2}. 
 We plotted the superfluid fraction obtained through
 the exact procedure (\ref{rs})  and the approximated one (\ref{boundapprox}), both for the configurations 
 with 20 and 100 particles.
 The agreement between the approximated curve and the exact one is good for large value of $\ell$
 while they start to differ when the localization parameter decreases, for values of the bound around $0.7$.
Unfortunately for the interesting values of $\ell$ the approximated calculation gives wrong results.
However, we find it useful, since it allows to estimate the typical scale of $\ell$ at which the bound starts
decreasing fast from 1 to 0 and we hope that it will be possible to improve it in the future, in order to be able to 
apply it to realistic cases.

\section{Conclusions}

The aim of this paper was to study Leggett's upper bound for amorphous quantum solids.
We showed that for quantum systems described by a hard sphere Jastrow wavefunction, the 
superfluid fraction must be smaller that $0.1\%$, which is consistent with a previous
investigation that found extremely small condensate fractions for this system~\cite{BCZ08}.
Moreover, the hard sphere result suggests that crystal and glass phase characterized by the same
Lindemann ratio  should have similar Leggett's upper bounds for the superfluid fraction.    

On this basis, we attempted to apply our results to glassy He$^4$~\cite{BPS1}. We found
that the upper bound for $\r_s$ is in general very close to the numerical results of Ref.~\cite{BPS1},
and at density $\r = 0.029$~$\AA$$^{-3}$ it is below.  One possible origin of this discrepancy could be that at 
such low density the life time of the metastable glassy state is too short, and the system is intrinsically
out of equilibrium; in that situation Leggett's bound is inapplicable, since it assumes that the reference
wave-function corresponds to a truly metastable state. 
Indeed we generically found from Path Integral Monte Carlo calculations that (at least if exchange is neglected) the 
system crystallizes very fast after the quench, which is consistent with a very short lifetime of the metastable glass.

Overall, our findings suggest two possible scenarios (not necessarly antithetic). 
(1) An amorphous stable glass has a superfluid fraction, 
not only a Leggett's upper bound, very similar to a defect-free crystal with the same 
Lindemann ratio. Since we know from 
experiments and simulations that this superfluid fraction 
is very small, or possibly zero, we are bound to conclude that
the glassy supersolid phase found in experiments do not correspond 
to a truly stable glass: the system is instead rapidly evolving 
out of equilibrium and, somehow, this enhances superfluidity. 
(2) Exchange promotes glassiness and whereas a  stable
glass phase cannot exist, because it has a very short life-time, 
a superglass can. This could be partially tested by 
comparing the stability of the glass phase in imaginary time 
simulations with and without exchange.    

It is worth to note that we neglected the role of small concentration of He$^3$ impurities
(of the order of few ppm) that has received a lot of attention in experiments~\cite{balibarreview}.
The reason is that we focused on a bulk glass phase of He$^4$, whose density
profile should be largely independent of such a small concentration of He$^3$ impurities.
It could be, however, that He$^3$ impurities affect the dynamical stability of the glass phase.
Based on the experience on classical systems, it is likely that in presence of a large concentration of impurities
crystallization will be avoided~\cite{KA95} and a long-lived quantum glass phase~\cite{QMCT,FSZ10} will be stable.
In this case, it should be very easy to measure the density profile and compute the Leggett bound using the 
procedure detailed above. However, it has been estimated that a concentration of at least $0.1\%$ of impurities 
is needed to stabilize the glass~\cite{oneper1000}. Therefore, the typical concentration of He$^3$ 
($\sim$ ppm) should not be enough to produce a sensible effect, unless some unexpected phenomenon related
to the quantum mechanical nature of the systems (e.g. exchange, as already discussed) becomes relevant.

\acknowledgments

We wish to thank S.~Baroni, M.~Boninsegni, G.~Carleo, S.~Moroni and L.~Reatto for very useful discussions.
FZ wishes to thank the Princeton Center for Theoretical Science for hospitality
during part of this work. 
This research was supported in part by the National Science Foundation under Grant No.~NSF~PHY05-51164.
Part of the numerical calculations have been performed on the cluster ``Titane'' of CEA-Saclay
under the grant GENCI 6418 (2010).


\appendix

\section{Details on the numerical procedure}
\label{app:A}

We define the Fourier transforms in the cubic box of side $L$ and volume $V=L^3$ as follows:
\beq
\r_{\vec q} = \frac1V  \int_V d\vec r \r(\vec r) e^{i \vec q \cdot \vec r} \ ,
\hskip20pt
\r(\vec r) = \sum_{\vec q}\r_{\vec q} e^{-i \vec q \cdot \vec r} \ ,
\eeq
where $\vec q = \frac{2 \pi}L (n_x,n_y,n_z)$, and each of the integers $n_i \in Z$,
and similarly 
\beq
\d\f_{\vec q} = \frac1V  \int_V d\vec r \d\f(\vec r) e^{i \vec q \cdot \vec r}  \ .
\eeq
Note that $\d\f_{\vec 0}$ is an irrelevant constant phase in the variational wavefunction
so we set it to zero. Finally,
\beq
\vec v_{\vec q} = \begin{cases}
\vec v_0 \hskip20pt \vec q = \vec 0 \ , \\
-i \vec q \f_{\vec q} \hskip20pt \vec q \neq \vec 0 \ . \\
\end{cases}
\eeq
which leads immediately to Eq.~(\ref{condiF}).

We performed the calculations for different values of the Lindemann parameter $\ell=\rho^{1/3}A^{1/2}$,
increasing the number of vectors $\vec q$ according to the spherical constraint $|\vec q| \leq q_{max}$,
until a reasonable convergence in the value of the bound (\ref{rs}) was achieved, at least for large values of $A$.
From Eq.~(\ref{rho}) one sees that for large $|\vec{q}|$ the corresponding component $\rho_{\vec{q}}$
is suppressed through the factor $e^{-A q^2/2}$. Thus, one needs to truncate the sum over  $\vec q$ 
at $q_{max} \sim 1/\sqrt{A}$, as higher terms will not contribute.
Unfortunately, for small $A$, this cut-off is too heavy in terms of computational time and 
we should use a lower one.
Still, considering small configurations and sufficiently large values of $A$, which nevertheless span the 
physical region of interest, we could reach a good convergence or keep the error under control. 
Note additionally that by increasing the number of vectors $\vec{q}$ in (\ref{rho}), 
the value found for the superfluid fraction 
monotonically decreases, as expected because of the variational property already discussed. This permits 
to preserve the nature of upper bound for Eq.~(\ref{rs}), despite the cut-off approximation.
Overall, we found that the better compromise was to set $q_{max} = 20\pi/L$.

In order to check the independence of the bound on the flow direction, we also compared the results
 obtained with the velocity $v_0$ along the $(1,0,0)$ direction to those along $(1,1,1)$ and we observed
  a negligible difference which is expected to vanish in the thermodynamic limit, because amorphous solids
 are statistically homogeneous on large scales.

We have also checked that the bound for the superfluid density almost does not fluctuate by considering
different amorphous configurations $\RR^\a$, $\a = 1, \cdots, \NN$, as it is expected since the superfluid
density is a macroscopic quantity. We computed the corresponding superfluid fraction $\r_s^\a$ 
and the average $\overline \rho_s = \sum_\a \r_s^\a / \NN$ for $10$ different configurations. The variance
of $\rho_s$ is very small. In this paper we presented results averaged over 10 realizations of $\mathcal{R}^{\alpha}$, 
 a larger statistics do not lead to appreciable differences.

\begin{figure}
\includegraphics[width=.5\textwidth]{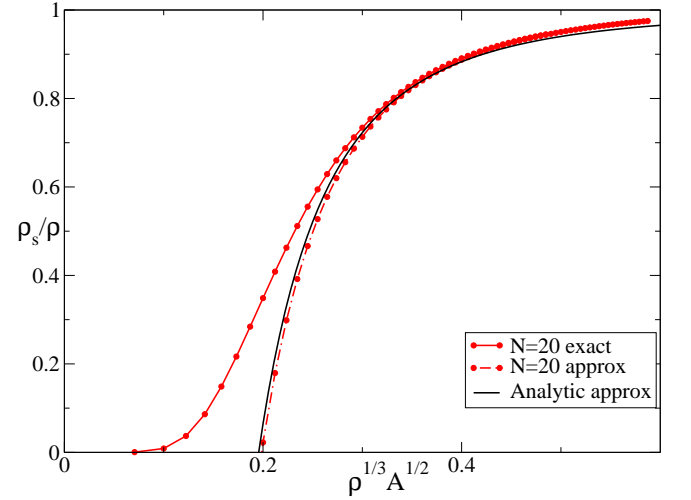}
\caption{
Result for $\overline{\r_s/\r}$ as a function of the localization parameter $\r^{1/3} A^{1/2}$,
where $\vec R_i$ are $N$ random points in $[0,L]^3$
with periodic boundary conditions.
}\label{random}
\end{figure}

Finally, as a check of our codes,
we repeated all the calculations on configurations of 20 particles occupying uncorrelated
uniformly random positions in the box, \ie where $\vec R_i$ are uniform and independent random variables
in $[0,L]^3$. In this case it is easy to show that $\wt g_q = \exp ( -A q^2)$.
Hence Eq.~(\ref{boundapprox}) becomes
\beq
\frac{\overline{ \rho_s}}{\r} = 1 -\frac2{3 (2 \pi)^2 \r} \int_0^\io dq \, q^2 \, e^{-A q^2}
= 1 -\frac1{24 \pi^{3/2} \, \r A^{3/2}} \ .
\eeq
In this case the values of the bound were more sensitive to the particular realization, so we took averages 
over 30 configurations.
For every value of the localization parameter, the superfluid fractions that we found were on average smaller, 
as reported in Figure \ref{random}.

\section{Large $A$ expansion}
\label{app:B}

For large $A$, we expect that the density becomes uniform. Hence, $\r_{\vec 0}\to \r$,
and $\r_{\vec q} \to 0$ for $\vec q \neq \vec 0$. We can use this to expand $i \d\f_{\vec q}$
systematically in powers of $\r_{\vec q}$.
We rewrite Eq.~(\ref{condiF}) as
\beq
\vec q \cdot \vec v_0 \r_{\vec q} = q^2 \r i\d\f_{\vec q} +  
\sum_{\vec p \neq \vec 0, \vec q} (\vec q \cdot \vec p) \r_{\vec q-\vec p} \, i\d\f_{\vec p} \ .
\eeq
We write $\d\f_{\vec q}= \d\f_{\vec q}^{(1)}+\d\f_{\vec q}^{(2)}+\cdots$ where the different terms
are of order $(\r_{\vec q})^k$.
At first order
\beq
i\d\f_{\vec q}^{(1)} = \frac{\vec q \cdot \vec v_0}{q^2 \r} \r_{\vec q} \ ,
\eeq
\begin{widetext}
at second order
\beq
i\d\f_{\vec q}^{(2)} = -\frac{1}{q^2 \r} 
\sum_{\vec p \neq \vec 0, \vec q} (\vec q \cdot \vec p) \r_{\vec q-\vec p} \, i\d\f_{\vec p}^{(1)} =
-\sum_{\vec p \neq \vec 0, \vec q}   \frac{(\vec q \cdot \vec p) (\vec p \cdot \vec v_0)}{p^2 q^2 \r^2}
\r_{\vec q-\vec p} \r_{\vec p} \ ,
\eeq
at third order
\beq
i\varphi_{\vec q}^{(3)}  = - \frac{1}{\vec q^2\rho} \sum_{\vec p\neq\vec0,\vec q} (\vec q\cdot\vec p)\rho_{\vec q-\vec p}i\varphi_{\vec p}^{(2)} 
 =\sum_{\vec p\neq\vec 0,\vec q}\sum_{\vec p'\neq\vec 0,\vec p}
\frac{(\vec q\cdot\vec p)(\vec p\cdot\vec p')(\vec p'\cdot\vec v_0)}{q^2 p^2 p'^2\rho^3}
\rho_{\vec q-\vec p}\rho_{\vec p-\vec p'}\rho_{\vec p'}
\eeq
from which we can guess the order $k$:
\beq
i\varphi_{\vec q}^{(k)} = (-1)^{k-1} \sum_{
\vec p_1\neq\vec 0,\vec q ; \ \vec p_{2} \neq\vec 0,\vec p_1 ; \
\cdots \ \vec p_{k-1} \neq\vec 0,\vec p_{k-2} }
\frac{(\vec q\cdot\vec p_1)(\vec p_1 \cdot\vec p_2) \cdots (\vec p_{k-1}\cdot\vec v_0)}
{ q^2 p_1^2\cdots p_{k-1}^2 \rho^k}
\rho_{\vec q-\vec p_1}\rho_{\vec p_1-\vec p_2} \cdots \rho_{\vec p_{k-2}-\vec p_{k-1}}\rho_{\vec p_{k-1}}
\eeq
and so on. Plugging this in Eq.~(\ref{rs}) we get
\beq\begin{split}
\frac{\r_s}{\r} =& 1 - 
\sum_{\vec q \neq \vec 0} \frac{(\vec v_0 \cdot \vec q)^2}{\r^2 v_0^2 q^2} \r_{\vec q}  \r_{-\vec q} 
+  \sum_{\vec q \neq \vec 0} 
\sum_{\vec p \neq \vec 0, \vec q}   
\frac{(\vec v_0 \cdot \vec q)(\vec q \cdot \vec p) (\vec p \cdot \vec v_0) }{q^2 p^2 v_0^2 \r^3 }
\r_{\vec q-\vec p} \r_{\vec p}\r_{-\vec q} \\
&- \sum_{\vec q\neq\vec0}\sum_{\vec p\neq\vec0,\vec q}\sum_{\vec p'\neq\vec0,\vec p}\frac{(\vec v_0\cdot\vec q)(\vec q\cdot\vec p)(\vec p\cdot\vec p')(\vec p'\cdot\vec v_0)}{ q^2  p^2  p'^2  v_0^2 \rho^4}\rho_{\vec q-\vec p}\rho_{\vec p-\vec p'}\rho_{\vec p'}\rho_{-\vec q} + \cdots
\ .
\end{split}
\eeq
While this expansion seems a simple strategy of solution of Eq.~(\ref{condiF}), it is very poorly convergent
and in practice it is not very helpful.
\end{widetext}

\end{document}